# Polarity and spin-orbit coupling induced large interfacial exchange coupling：an asymmetric charge transfer in iridate-manganite heterostructure

Tao Yu [a,†], Bei Deng [a,†], Liang Zhou[a], Pingbo Chen[a], Qiying Liu[a], Xingkun Ning[b], Jingtian Zhou[c], Zhiping Bian[c], Zhenlin Luo[c], Chunyin Qiu[d], Xingqing Shi[a,*], Hongtao He[a,*]

[a] *Department of Physics, Southern University of Science and Technology, Shenzhen, 518055, China*

[b] *Hebei Key Lab of Optic-electronic Information and Materials, The College of Physics Science and Technology, Hebei University, Baoding 071002, China*

[c] *National Synchrotron Radiation Laboratory and CAS Key Laboratory of Materials for Energy Conversion, University of Science and Technology of China, Hefei, Anhui 230026, China*

[d] *Key Laboratory of Artificial Micro- and Nano-structures of Ministry of Education and School of Physics and Technology, Wuhan University, Wuhan 430072, China*

**Abstract**

Charge transfer is of particular importance in manipulating the interface physics in transition-metal oxide heterostructures. In this work, we have fabricated epitaxial bilayers composed of polar 3$d$ LaMnO$_3$ and nonpolar 5$d$ SrIrO$_3$. Systematic magnetic measurements reveal an unexpectedly large exchange bias effect in the bilayer, together with a dramatic enhancement of the coercivity of LaMnO$_3$. Based on first-principles calculations and x-ray absorption spectroscopy measurements, such a strong interfacial magnetic coupling is found closely associated with the polar nature of LaMnO$_3$ and the strong spin-orbit interaction in SrIrO$_3$, which collectively drives an asymmetric interfacial charge transfer and leads to the emergence of an interfacial spin glass state. Our study provides new insight into the charge transfer in transition-metal oxide heterostructures and offer a novel means to tune the interfacial exchange coupling for a variety of device applications.

*Corresponding authors at: Department of Physics, Southern University of Science and Technology, Shenzhen, 518055, China.
*E-mail address*: heht@sustc.edu.cn (H.T.H); shixq@sustc.edu.cn (X.Q.S)
†These authors contributed equally to this work

## 1. Introduction

The synthesis of dissimilar complex oxide heterostructures is currently one of the hottest areas in the design of novel functional materials. Due to the strong interplay among charge, spin, orbital, and lattice degrees of freedom, the interface between different transition-metal oxides (TMOs) in artificially layered heterostructures exhibits many exotic physical properties that are absent in the constituent bulk materials.[1-5] Among many factors, the interfacial charge transfer has been identified as an effective knob to tune the interface physics, such as mediating the interfacial magnetic coupling in $YBa_2Cu_3O_7/La_{0.66}Ca_{0.33}MnO_3$[8,9] and $La_{0.75}Sr_{0.25}MnO_3/LaNiO_3$[10] heterostructures. Generally, the charge transfer can be driven by the work function differences between the contact TMOs [11] or even by their polarity discontinuity[1,12-14]. The most prominent model of polarity discontinuity is the emergence of two-dimension electron gas at the interface of two band insulators: polar $LaAlO_3$ and nonpolar $SrTiO_3$[1,12,13]. Novel phenomena, such as the insulator-to-metal transition and magnetism emerged at the interface of such heterostructures have been reported. Therefore, by manipulating the polarity discontinuity at the interface, one can effectively modulate the interfacial physical properties of TMO heterostructures[15].

Recently, the 5*d* iridium oxides, in which the large spin-orbit coupling (SOC) and on-site Coulomb interaction exhibit a comparable energy scale, have attracted considerable attention due to theoretical predictions of unconventional phases like superconductivity, topological Mott insulator, and Weyl semi-metals[16-19]. But so far, only a few exotic phenomena arising from the strong interfacial coupling in perovskite $SrIrO_3$-based heterostructures have been reported[20-24]. One typical example is the observation of ferromagnetic ground state and strong anomalous Hall effect at the interface between antiferromagnetic (AFM) $SrMnO_3$ and paramagnetic (PM)

SrIrO$_3$[2]. Despite the nonpolar interface and nearly identical work functions, an interfacial charge transfer giving rise to the observed phenomena can still occur due to the coupling of molecular *d* orbitals and strong SOC in SIO[25]. Up to now, the combined effect of polarity discontinuity and strong SOC on the interface physics of heterostructures has been rarely studied in the literature, which may thus offer challenges as well as opportunities in manipulating the interfacial properties in TMO heterostructures for a wide range of potential applications.

Here, we have investigated the charge transfer at the interface between nonpolar *5d* SrIrO$_3$ (SIO) and polar *3d* LaMnO$_3$ (LMO) epitaxially grown on (001) SrTiO$_3$ single-crystalline substrates. The stochiometric bulk LMO with electronic configuration $t_{2g}^3 e_g^1$ is an A-type antiferromagnetic band insulator. In the form of epitaxial thin films, however, LMO often exhibits ferromagnetism (FM) with a Curie temperature ($T_C$) of ~150 K. Although the origin of its ferromagnetism is still under debate, some possible mechanisms such as vacancies, epitaxial strain, and the charge transfer resulting from its polar nature have been raised to understand the emergent ferromagnetism[26-28]. Meanwhile, the SrIrO$_3$ epitaxial layer with strong SOC is paramagnetic. Owing to the interplay between the polarity-induced charge transfer and the charge transfer driven by molecular orbital coupling (MOC) and SOC, a strong interfacial magnetic coupling appears in the heterostructure, which manifests itself in the dramatic enhancement of coercivity and the observation of the largest exchange-bias effect observed so far in PM/FM systems[29,30]. In addition, first-principles calculations have been performed to clarify the charge transfer at the interface between SrIrO$_3$ and LaMnO$_3$. The calculated results not only explain the experimental observation but are further verified by the x-ray absorption spectroscopy (XAS) measurements.

**II．Experimental Details**

The epitaxial bilayers composed of SIO and LMO were deposited on SrTiO$_3$ (001) substrates by pulsed-laser deposition (PLD) system. The energy fluence of the laser is approximately 1.2 J/cm$^2$ with a repetition rate of 2Hz. The depositions were performed in an oxygen pressure of 0.18 mbar and at a substrate temperature of 700℃. After growth, the samples were in situ annealed at 700 ℃ in a pressure of 0.6 bar pure O$_2$ for 0.5 h and were then slowly cooled down to room temperature at a rate of 5 ℃/min to remove possible oxygen vacancies. The microstructures of these samples were investigated by X-ray diffraction (XRD) (Rigaku, Smartlab, Cu $K\alpha$ radiation, 0.15406 nm) and transmission electron microscopy (TEM, Titan TM Themis G2 60-300). Magnetic measurements were performed in a superconducting quantum interference device (SQUID) magnetometer (Quantum Design MPMS3), with the field applied parallel to the film plane. Meanwhile, the thin bilayers were also grown under the same conditions in order to investigate the valence state of Mn cations near the interface by the x-ray absorption spectroscopy (XAS) measurements. First-principles calculation was performed to probe the charge transfer between SrIrO$_3$ and LaMnO$_3$ in the bilayers. The first principles calculations were carried out within density functional theory as implemented in the VASP code[31]. The lattice geometries and electronic spin structures of SrIrO$_3$, LaMnO$_3$ and their heterostructures were treated with the LSDA+U functional, while the Coulomb repulsive potential U term was parameterized with 2 eV for the 5$d$ states of Ir and 3$d$ states of Mn, and 6 eV for the 4$f$ orbitals of La, respectively. PAW pseudopotentials[32] for Sr, La, Ir, Mn and O were used with a plane-wave basis-set cutoff energy of 450 eV. A 5×5×1 Monkhorst-Pack mesh was used for the k-point samplings of (2×2) heterostructure slabs in their Brillouin zones. In obtaining the lattice geometry, all the atoms in the slab were allowed to relax until the calculated Hellmann-Feynman forces are less than 3×10−3 eV/Å. Spin-orbit coupling was taken into account in the calculations except mentioned otherwise.

**III. Results and Discussion**

The XRD patterns around the (001) reflection for the LMO and SIO single layers, and SIO/LMO bilayer heterostructure in Fig. 1 show a (001)-preferred orientation, indicating the sequential epitaxy of the SIO and LMO layers on the STO substrate. No secondary phases are detected in these samples. The high-resolution TEM (HRTEM) image of the SIO/LMO bilayer sample demonstrates the high crystalline quality of the bilayer. It is noted that the interface between SIO and LMO is clear and well-defined with the film thicknesses of about 25 nm and 8 nm, respectively. Meanwhile, there are no amorphous layer and detectable interdiffusion throughout the cross section of the bilayer. These results thus collectively verify both the SIO and LMO layers have grown epitaxially on the STO substrate with the high crystalline quality.

The hysteresis loops of the SIO/LMO bilayer, measured at 2 K after zero-field cooled (ZFC) and field-cooled (FC) in a field of +/- 4000 Oe from the room temperature, are shown in Fig. 2 (a). For comparison, the ZFC hysteresis loops of the SIO and LMO single layers are also given in the inset of Fig. 2 (a). An obvious shift along the magnetic-field axis has been observed in the bilayer, indicating the presence of an unexpected exchange-bias (EB) effect in the PM/FM bilayer. The shift in the hysteresis loop may be quantified through the exchange-bias field $H_E=|H_L+H_R|/2$, whereas the coercivity is calculated by $H_C=|H_L-H_R|/2$, and the $H_L$ and $H_R$ are the lower and higher field value, respectively, where the average film magnetization becomes zero. A large exchange field $H_E$ of 432 Oe is observed in the SIO/LMO bilayer. To the best of our knowledge, this is the largest value of $H_E$ reported so far in PM/FM systems[10, 29, 30]. Moreover, the coercivity $H_C$ dramatically increases from 315 Oe for the LMO single layer to 1722 Oe for the bilayer. Obviously, the SIO single layer is typically paramagnetic and could not contribute any magnetic moments. Meanwhile, the LMO single layer does not exhibit any EB effect, the observed EB effect in the bilayer should unambiguously arise from the strong interfacial coupling. The temperature dependencies of $H_E$ and $H_C$ for the SIO/LMO bilayer after FC in 4000 Oe from the room temperature are shown in Fig. 2 (b). It is noted that the $H_E$ decreases with increasing the temperature, eventually vanishing at the so-called blocking temperature $T_B$ of about 60 K. On the other hand, the $H_C$ exhibits a similar trend that with increasing temperature, $H_C$ monotonically decays. Besides this, it is also found that the temperature dependencies of $H_E$ and $H_C$ follow the phenomenological formula $H_E(T) = H_E^0 \exp(-T/T_1)$ and $H_C(T) = H_C^0 \exp(-T/T_2)$, respectively, where $H_E^0$ and $H_C(T)$ are the extrapolations of $H_E$ and $H_C$ at 0 K, and $T_1$ and $T_2$ are constants. As shown in Fig. 2 (b), both $H_E$ and $H_C$ exponentially decay with increasing temperature. Similar experimental results were also observed in perovskite manganite such as FM/AFM $La_{0.67}Ca_{0.33}MnO_3/La_{0.33}Ca_{0.67}MnO_3$ multilayers[33], FM/PM $La_{0.75}Sr_{0.25}MnO_3(LSMO)/LaNiO_3$[10], and $La_{0.7}Sr_{0.3}MnO_3/SrMnO_3$[34] bilayers, where the existence of the frustration due to competing magnetic interactions is known to lead to an exponential decay of $H_E$ and $H_C$.

In order to investigate the origin of the unexpectedly large EB effect in our PM/FM bilayer, we have measured the magnetization as a function of temperature $M(T)$ for the SIO/LMO bilayer as well as the LMO single layer in an in-plane magnetic field of 100 Oe, as shown in Fig 2 (c). For the single layer, both the ZFC and FC $M(T)$ curves coincide with each other and the magnetization decreases monotonically with increasing temperatures, with the FM to PM transition occurring around 150 K. However, the bilayer sample exhibits two striking features: a bifurcation between the $M_{ZFC}(T)$ and $M_{FC}(T)$ curves below an irreversibility temperature ($T_{irr}$) and a maximum of magnetization appearing around $T_p$ in the $M_{ZFC}(T)$ curve. These phenomena collectively suggest that some sorts of frozen states exist at the interface of our bilayer[35,36]. It is also noted that both $T_p$ and $T_{irr}$ are reduced when the applied field is enhanced, indicating that the frozen state is suppressed by a strong magnetic field, as shown in Fig. 2 (d). Furthermore, the field dependence of $T_{irr}$ is found to follow the so-called Almeida-Thouless line which is characteristic of the spin glass (SG)[37]: $H(T_{irr})/\Delta J \propto (1 - T_{irr}/T_F)^{3/2}$, where $T_F$ is the zero-field SG freezing temperature and $\Delta J$ the width of the distribution of exchange interactions. The inset of Fig. 2 (d) shows the linear fit of $H^{2/3}$ as a function of $T_{irr}$. Therefore, an interfacial SG state is most likely to emerge in our bilayer, with a freezing temperature $T_F \sim 84$ K. It is this interfacial SG state that couples to the FM LMO layer and results in the observed EB effect shown in Fig. 2 (a), consistent with previous studies where unexpected EB effects were also observed in $LaNiO_3$-based heterostructures[10,29,30]. Also in these studies, the emergence of interfacial SG was ascribed to the interfacial charge transfer. The variation in the valence state due to the interfacial charge transfer would change the interfacial magnetic behavior. An interfacial SG state would thus be formed as a result of competing magnetic interactions with different strengths and signs among the interfacial Mn and Ni cations[10]. But in comparison with the LaNiO$_3$-based heterostructures, the charge transfer in our bilayer would be more complicated, since we need to consider not only the polar nature of LMO, but also the strong SOC in SIO.

On one hand, based on the first-principles calculations and combined with the XAS characterization, it has been shown that owing to the spin-orbit coupling effect, a charge transfer can indeed take place in the SrMnO$_3$-SrIrO$_3$ interface[2, 25] despite their negligible work function difference and the charge non-polarity nature. As depicted in Fig. 3 (a), under the $O_h$ crystal filed of Mn-O$_6$ octahedron and by coupling with the O 2$p$ states, the Mn 3$d$ states will split into lower-lying $t_{2g}$ manifold and higher-lying $e_g$ doublet. The resultant $e_g$ states of Mn are favorable to couple with the $e_g$ states of Ir, which are higher above the Fermi energy, giving rise to a lower-lying molecular orbital (with bonding character), and a higher-lying molecular orbital above the Fermi energy (with anti-bonding character). Due to the strong spin-orbit interaction in SIO, the $t_{2g}$ states of Ir are further split into $J$=3/2 and $J$=1/2 states, with the latter higher in energy. This offers a tendency for the electrons in the higher-lying $J$=1/2 states transferring into the lower-lying $e_g$ molecular orbitals, as shown in Fig. 3 (b). In our case, however, the Mn ion in LMO nominally takes a valency of +3 and, hence, the $e_g$ state of Mn$^{3+}$ happens to be half-occupied, very different from the fully unoccupied $e_g$ state of Mn$^{4+}$ in the SMO case. Therefore, by further coupling with the O 2$p$ states, the Mn$^{3+}$ $e_g$ states should be lower in energy than the Mn$^{4+}$ ones, leading to a larger energy difference with respect to the $J$=1/2 states of Ir. This scenario indicates that the interfacial-charge-transfer-strengthened exchange coupling would be more favorable in our SIO/LMO bilayer.

On the other hand, the interfacial charge transfer is to a large extent influenced by the polarity discontinuity in heterostructures composed of polar and nonpolar layers. Since LMO is a polar material consisting of alternatively charged LaO$^{+1}$ and MnO$_2^{-1}$ planes, whereas SIO contains charge neutral SrO and IrO$_2$ planes, a polar discontinuity appears at the interface of our (001) SIO/LMO heterostructures. Note that there could be two different interface configurations, *i.e.*, the SrO/IrO$_2$/LaO/MnO$_2$ interface or the IrO$_2$/SrO/MnO$_2$/LaO interface, as depicted in Fig. 4 (a) and (b), respectively. In order to eliminate the polar discontinuity, the formally charged LaO layer will spontaneously donate more or less half of an electron to the adjacent IrO$_2$ layer upon the formation of the SrO/IrO$_2$/LaO/MnO$_2$ geometry (see Fig. 4 (a)). On the contrary, the SrO layer will donate more or less half of an electron to the neighboring MnO$_2$ layer upon the formation of the IrO$_2$/SrO/MnO$_2$/LaO interface (see Fig. 4 (b)). If there is no defect involved in the SrO layer, the resulting positive charge will be accommodated by the electronic reconstruction of Ir ions in the adjacent IrO$_2$ layer,[15] *i.e.*, a charge transfer from the IrO$_2$ to SrO layer, as indicated in Fig. 4 (b) by the red arrow. In the first case, the polarity-driven long-range charge transfer will not change the charge population of the interfacial Mn, but it will certainly increase the charge population of the interfacial Ir, in another words, the orbital population of the $J=1/2$ states at the interface. Then, the charge transfer occurring from the $J=1/2$ states of Ir to the lower-lying $e_g$ states of Mn due to the $d$ orbital coupling favored by SOC will reasonably increase the charge population of the interfacial Mn atoms, as shown in Fig. 4 (a) by the green arrow. Therefore, the interfacial charge transfer from the $J=1/2$ states to the $e_g$ states will be enhanced as a combined effect of the polarity-induced charge transfer and the charge transfer driven by molecular orbital coupling (MOC) and SOC. In the second case, however, the polarity-induced charge transfer does not change the charge population of the interfacial Mn atoms but lowers the charge population of the interfacial Ir, which in turn reduces the orbital occupancy of the $J=1/2$ states and, as well, weakens the charge transfer to the $e_g$ states of Mn on the basis of MOC and SOC pictures. The above discussion not only sheds new lights on the long-standing problem of charge-transfer mechanism in TMO heterostructures, but also offers a novel means to tune the interfacial exchange coupling for a variety of device applications.

In order to determine the real interface configuration in our SIO/LMO bilayers, we have performed XAS measurements on thin SIO/LMO bilayers which were grown under the same conditions. We have collected the Mn $L$-edge XAS spectra for the bilayers, with the single LMO layer as a reference, as shown in Fig. 5, which can to a good characterization reflect the charge state variation of the overall Mn atoms (not the interfacial Mn only). The most striking finding here is the shift of the XAS peak position of the Mn $L_3$ edge to higher energy in bilayer samples, indicating that the Mn oxidation state is enhanced with respect to the LMO single layer. Our XAS results convincingly confirm the overall long-range charge transfer occurs from the LMO to the interfacial SIO layer. But for the $IrO_2$/SrO/$MnO_2$/LaO configuration, the overall charge transfer is from the interfacial SIO layer to LMO layers (see Fig. 4 (b)). Therefore, we can safely conclude that the dominant interface configuration in our bilayers is SrO/$IrO_2$/LaO/$MnO_2$, as depicted in Fig. 4 (a).

To verify the above picture, we have also performed first-principles calculations based on the SrO/$IrO_2$/LaO/$MnO_2$ configuration. The calculated density of states (DOS) are shown in Fig. 6. As expected, the majority spin states of Mn are split in to the $t_{2g}$ and $e_g$ manifolds. The $e_g$ states consisting of $d_{x2-y2}$ and $d_{z2}$ orbitals have a strong coupling with the higher-lying $e_g$ states of Ir, especially for the $d_{x2-y2}$ orbitals, which exhibit a broadening character. Besides this, the coupling between Mn and Ir $t_{2g}$ states also pushes the Ir $t_{2g}$ orbital shifting upward. Therefore, the Ir $t_{2g}$ states are higher in energy than the $e_g$ bonding states of Mn, giving rise to a pathway for the charge transfer between these two states (see Fig. 6 (a) & (b)). Such a charge transfer can be further enhanced by SOC in the interfacial SIO layer. As shown in Fig. 6 (c), the Ir $t_{2g}$ states are further split into the higher-lying $J=1/2$ states and lower-lying $J=3/2$ states. The rise of the $J=1/2$ states in energy will thus promote more electrons flowing from the Ir $J=1/2$ states to the Mn $e_g$ bonding states at SIO-LMO interface, whereby largely enhance the ferromagnetic exchange coupling between the two constitute parts.

Since the magnitude change of magnetic moments can to a good characterization reflect the charge transfer across the interface quantitively[25], we have also calculated the magnetic moment of the Mn and Ir atoms at the interface optionally effected by the SOC effect. We can explicitly find that the magnetic moment of Mn is enhanced by ~3% (from 3.8 $\mu B$ to 3.91 $\mu B$) in the SrO/$IrO_2$/LaO/$MnO_2$ interface with respect to the bulk configuration. After inclusion of the SOC effect, this value is further increased to ~4% (from 3.8 $\mu$B to 3.95 $\mu$B). On the contrary, the magnetic moment of Ir is reduced by ~61% (from 0.62 µB to 0.38µB, because the spin moment of Ir is dominated by the spin-down channel, as is revealed in Fig. 6 (b), so a decrease of spin moment corresponds to process of charge gain) with respect to the bulk configuration, but increased from 0.38µB to 0.42 µB after considering the SOC. Accordingly, the calculated charge transfer to the interfacial Mn ions is about 0.11 e$^-$ per (1×1) unit cell, which is enhanced to about 0.15 e$^-$ per(1×1) unit cell as long as the SOC effect is taken into account, consistent with the above picture.

It should be noted here that the magnetic moment change of the interfacial Mn is not equal to that of Ir ions. This is due to the fact that while the charge transfer on the interfacial Mn ions is mainly driven by the MOC and SOC, the charge transfer on the interfacial Ir can be further affected by the interfacial polarity discontinuity, *i.e*, the interplay between the polarity-driven charge transfer and the charge transfer induced by MOC and SOC collectively determines the magnetic moments of the interfacial Ir ions. The amount of charge all over the MnO layers long-ranged transferred to the interfacial IrO layer is more or less half of an electron [here ~0.35 e$^-$ (0.24 e$^-$+0.11e$^-$) for the interfacial Ir] due to the polarity discontinuity at the SrO/IrO$_2$/LaO/MnO$_2$ interface. Whereas the calculated charge transfer on the interfacial Mn ions is only about 0.15 e$^-$ per(1×1) unit cell as long as the SOC effect is taken into account. Therefore, the interfacial charge transfer in SIO-LMO heterostructure exhibits a typical asymmetric character in respect of the conventional interfacial charge transfer, which should be responsible for the novel interfacial magnetic behaviors as observed in our study. Also, it is important to note that the average Mn oxidation state in the bilayer sample should be even higher than that of the LMO single layer, as indeed confirmed by the XAS measurement above in Fig. 5. In our bilayer case, the charge transfer occurring at the interface between nonpolar SIO and polar LMO leads the interfacial Mn and Ir cations appear to take mixed valence states. The competing magnetic interactions of different strengths and signs among the interfacial Mn and Ir ions may thus give rise to an interfacial SG state, which pins the ferromagnetic LMO layer below the blocking temperature and results in the largest EB effect observed so far in PM/FM systems.

## IV．Conclusion

In summary, we have systematically studied the interfacial magnetic coupling in nonpolar 5$d$ SrIrO$_3$ (paramagnetic) polar and 3$d$ LaMnO$_3$ (ferromagnetic) bilayer grown epitaxially by PLD. Due to the strong interfacial magnetic coupling, an unexpectedly large EB effect and dramatic enhancement of the coercivity have been observed in the bilayer. The XAS experiment results and first-principle calculations show the charge transfer occurring at the interface of our bilayer, owing to the interplay between interfacial charge polarity discontinuity and the orbital coupling accompanied with SOC. The competing magnetic interactions of interfacial Mn and Ir cations with the mixed valence states resulting from the charge transfer give rise to the emergence of the interfacial SG state, which couples to the ferromagnetic LMO layer and induces the unexpected EB effect.

## Acknowledgement

This work was supported by the National Natural Science Foundation of China (No. 11574129), the National Key Research and Development Program of China (No. 2016YFA0301703), Technology and Innovation Commission of Shenzhen Municipality (Nos. KQJSCX20170727090712763 and JCYJ20170817105007999) and the Center for Computational Science and Engineering of Southern University of Science and Technology.

## References

1. A. Ohtomo, H.Y. Hwang, Nat. 427 (2004) 423.
2. J. Nichols, X. Gao, S. Lee, T.L. Meyer, J.W. Freeland, V. Lauter, D. Yi, J. Liu, D. Haskel, J.R. Petrie, E.J. Guo, A. Herklotz, D. Lee, T.Z.Ward, G. Eres, M.R. Fitzsimmons, H.N. Lee, Nat. Comm.7 (2016) 12721.
3. K.S. Takahashi, M. Kawasaki, Y. Tokura, Appl. Phys. Lett. 79, (2001) 1324.
4. M. Gilbert, P. Zubko, R. Scherwitzl, J. Iniguez, J.M. Triscone, Nat. Mater. 11

(2012) 195.

5. J. Hoffman, I. C. Tung, B. B. Nelson-Cheeseman, M. Liu, J. W. Freeland, A. Bhattacharya, Phys. Rev. B 88 (2013) 144411.

6. H. Chen, A.J. Millis, and C. A. Marianetti, Phys. Rev. Lett. 111 (2013) 116403.

7. T.S. Santos, S.J. May, J. Robertson and A. Bhattacharya, Phys. Rev. B 80 (2009) 155114.

8. P. Przyslupski, I. Komissarov, W. Paszkowicz, P. Dluzewski, R. Minikayev, and M. Sawicki, Phys.Rev. B 69 (2004) 134428.

9. N. Haberkorn, J. Guimpel, M. Sirena, L. B. Steren, W. Saldarriaga, E. Baca, and M. E. Gómez, Appl. Phys. Lett. 84 (2004) 3927.

10. J.C. Rojas Sánchez, B Nelson-Cheeseman, M. Granada, E. Arenholz, L.B. Steren, Phys. Rev. B 85 (2012) 094427.

11. S. Mubeen, J. Lee, and W. Lee et al., ACS. Nano. 8, 6066 (2014).

12. N. Reyren, S. Thiel, A. D. Caviglia, L. F. Kourkoutis, G. Hammerl, C. Richter, C. W. Schneider, T. Kopp, A. S. Rüetschi, D. Jaccard, M. Gabay, D. A. Muller, J. M. Triscone, and J. Mannhart, Sci. 317 (2007) 1196.

13. A. Brinkman, M. Huijben, M. van Zalk, J. Huijben, U. Zeitler, J. C. Maan, W. G. van der Wiel, G. Rijnders, D. H. Blank, and H. Hilgenkamp, Nat. Mater. 6 (2007) 493.

14. X. R. Wang, C. J. Li, W. M. Lü, T. R. Paudel, D. P. Leusink, M. Hoek, N. Poccia, A. Vailionis, T. Venkatesan, J. M. D. Coey, E. Y. Tsymbal, Ariando, H. Hilgenkamp, Sci.349 (2015) 716.

15. N. Nakagawa, H. Y. Hwang, and D. A. Muller, Nat. Mater. 5 (2006) 204.

16. S.J. Moon, H. Jin, K.W. Kim, W.S. Choi, Y.S. Lee, J. Yu, G. Cao, A. Sumi, H. Funakubo, C. Bernhard, T.W. Noh, Phys. Rev. Lett. 101 (2008) 226402.

17. D. Xiao , W. Zhu, Y. Ran, N. Nagaosa and S. Okamoto, Nat. Commun. 2 (2011) 596.

18. Y. K. Kim, O. Krupin, J. D. Denlinger, A. Bostwick, E. Rotenberg, Q. Zhao, J. F. Mitchell, J. W. Allen, and B. J. Kim, Sci., 345 (2014) 187.

19. J. W. Kim, Y. Choi, J. Kim, J. F. Mitchell, G. Jackeli, M. Daghofer, J. van den

Brink, G. Khaliullin, and B. J. Kim, Phys. Rev. Lett. 109 (2012) 037204.

20. J. Matsuno, N. Ogawa, K. Yasuda, F. Kagawa, W. Koshibae, N. Nagaosa, Y. Tokura, M. Kawasaki, Sci. Adv. 2, (2016) e1600304.

21. B. Pang, L.Y. Zhang, Y.B. Chen, J. Zhou, S.H. Yao, S.T. Zhang, Y.F. Chen, ACS Appl. Mater. Interfaces, 9 (2017) 3201.

22. D. Yi, C.L. Flint, P.P. Balakrishnan, K. Mahalingam, B. Urwin, A. Vailionis, A.T. N'Diaye, P. Shafer, E. Arenholz, Y. Choi, K.H. Stone, J.H. Chu, B.M. Howe, J. Liu, I.R. Fisher, Y. Suzuki, Phys. Rev. Lett. 119 (2017) 077201.

23. L. Hao, D. Meyers, C. Frederick, G. Fabbris, J.Y. Yang, N. Traynor, L. Horak, D. Kriegner, Y. Choi, J.W. Kim, D. Haskel, P.J. Ryan, M.P.M. Dean, J. Liu, Phys. Rev. Lett. 119 (2017) 027204.

24. J. Matsuno, K. Ihara, S. Yamamura, H. Wadati, K. Ishii, V.V. Shankar, Hae-Young Kee, H. Takagi, Phys. Rev. Lett. 114 (2015) 247209.

25. S. Okamoto, J. Nichols, C. Sohn, S.Y. Kim, T.W. Noh, H.N. Lee, Nano Lett. 17 (2017) 2126.

26. A. Gupta, T. R. McGuire, P. R. Duncombe, M. Rupp, J. Z. Sun, W. J. Gallagher, and Gang Xiao, Appl. Phys. Lett. 67 (1995) 3494.

27. S. Dong, R. Yu, S. Yunoki, G. Alvarez, J.M. Liu, and E. Dagotto, Phys. Rev. B 78 (2008) 201102.

28. X.R. Wang, C.J. Li, W.M. Lü, T.R. Paudel, D.P. Leusink, M. Hoek, N. Poccia, A. Vailionis, T. Venkatesan, J.M.D. Coey, E.Y. Tsymbal, Ariando, and H. Hilgenkamp, Sci.349 (2015) 6249.

29. M. Gilbert, P. Zubko, R. Scherwitzl, J. Iniguez, J.M. Triscone, Nat. Mater. 11 (2012) 195.

30. X.K. Ning, Z.J. Wang, Z.D. Zhang, Sci. Rep. 5 (2015) 8064.

31. G. Kresse, J. Furthmüller, Phys. Rev. B 54 (1996) 11169.

32. G. Kresse, D. Joubert, Phys. Rev. B 59 (1999) 1758.

33. I. Panagiotopoulos, C. Christides, M.Pissas and D. Niarchos, Phys. Rev. B 60 (1999) 485.

34. J.F. Ding, O.I. Lebedev, S. Turner, Y.F. Tian, W.J. Hu, J.W. Seo, C.

Panagopoulos, W. Prellier, G. van Tendeloo, T. Wu, Phys. Rev. B 87 (2013) 054428.

35. S. Karmakar, T. Taran, E. Bose, B.K. Chaudhuri, C.P. Sun, C.L. Huang and H.D. Yang, Phys. Rev. B 77 (2008) 144409.

36. X.H. Huang, J.F. Ding, Z.L. Jiang, Y.W. Yin, Q.X. Yu, and X.G. Li, J. Appl. Phys. 106 (2009) 083904.

37. K. Binder, A.P. Young, Rev. Mod. Phy. 58 (1986) 801.

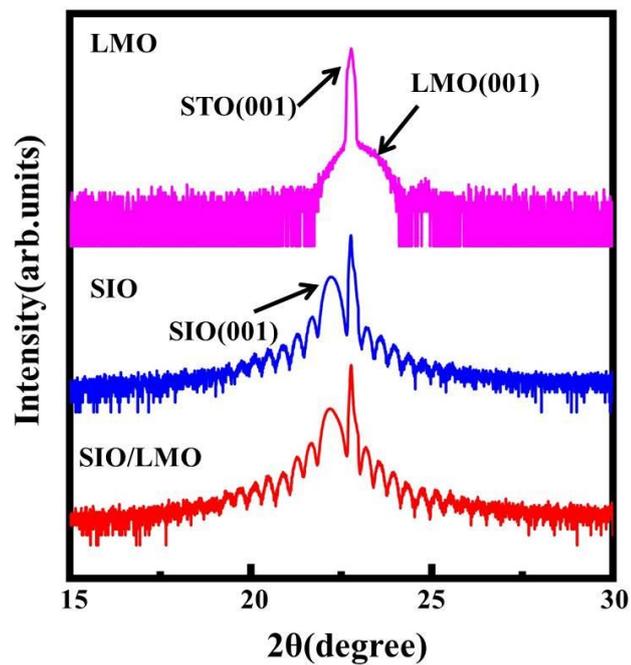

Figure 1. XRD scans around the (001) Bragg reflection for the SIO/LMO bilayer and the SIO and LMO single layers for reference.

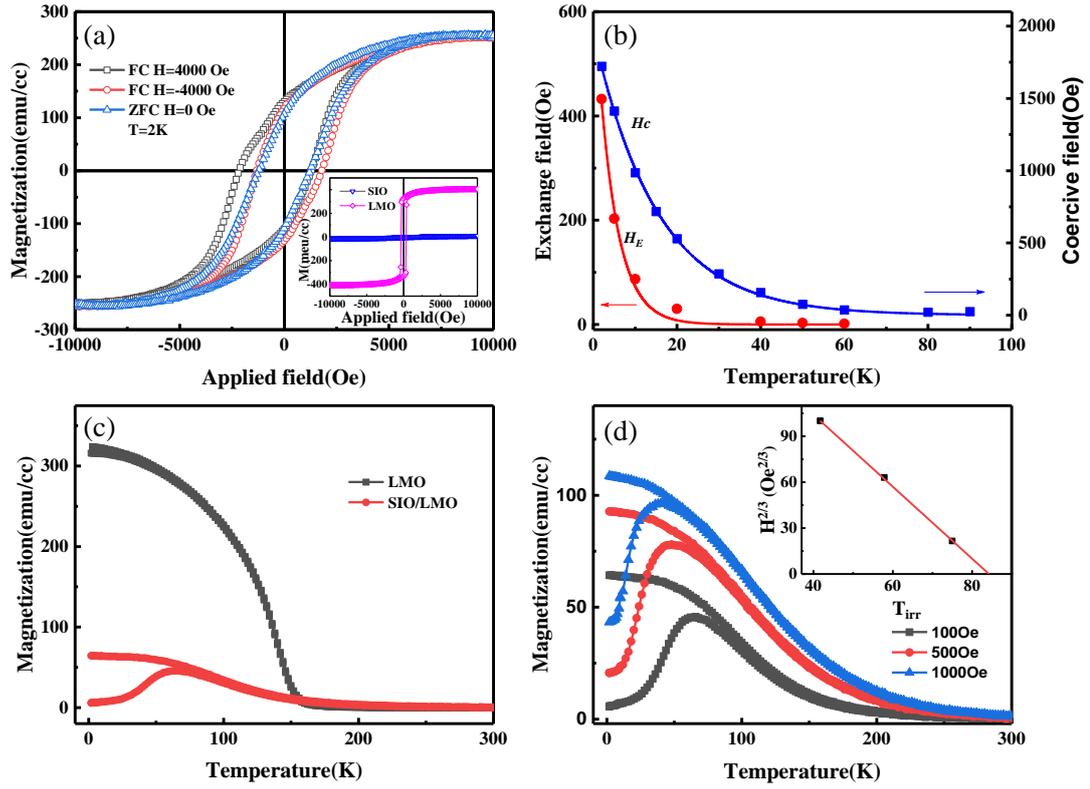

Figure 2. The magnetic hysteresis loops of the SIO/LMO bilayer at 2 K after zero-field cooling (ZFC) and field-cooled (FC) in a field of +/- 4000 Oe. (b) Temperature dependence of the exchange bias field $H_E$ and coercivity $H_C$. (c) Temperature dependence of magnetization measured in an magnetic field of 100 Oe for the bilayer. (d) Temperature dependence of magnetization measured under different magnetic fields(100、500、1000 and 2000 Oe). Inset: Corresponding plot of $H^{2/3}$ vs $T_{irr}$ and the red line is the fitting to equation: $H(T_{irr})/\Delta J \propto (1 - T_{irr}/T_F)^{3/2}$.

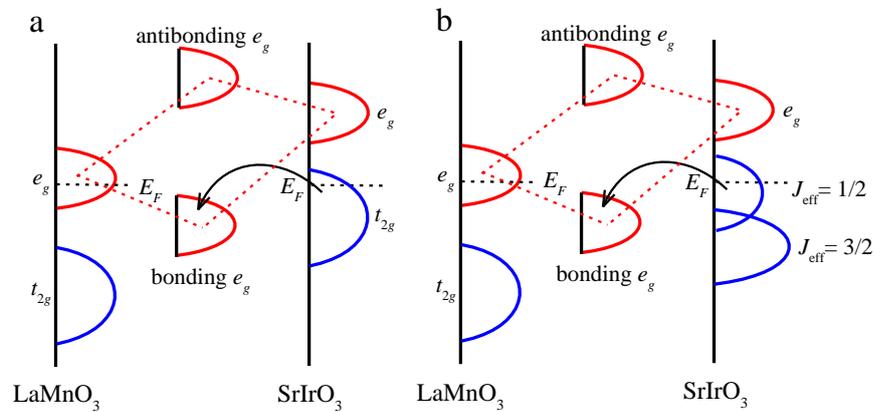

Figure 3. Sketch map of (a) charge transfer at LaMnO$_3$–SrIrO$_3$ interface due to the crystal-filed splitting and molecular orbital coupling between e$_g$ states of Mn and Ir, and (b) the charge transfer is enhanced by spin-orbit coupling of Ir t$_{2g}$ states.

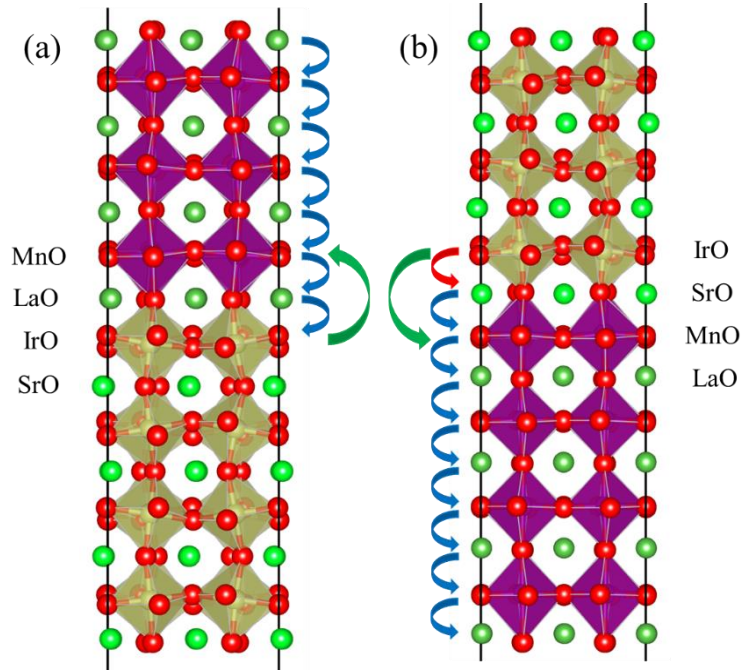

Figure 4. Illustrated graphs of the charge transfer in LaMnO$_3$-SrIrO$_3$ interface. The blue arrows show the charge transfer due to the interface polarity, while the red arrow display the charge transfer caused by the electronic reconstruction and the green ones denote the charge transfer by molecular orbital coupling and spin-orbit coupling as illustrated in Figure 3.

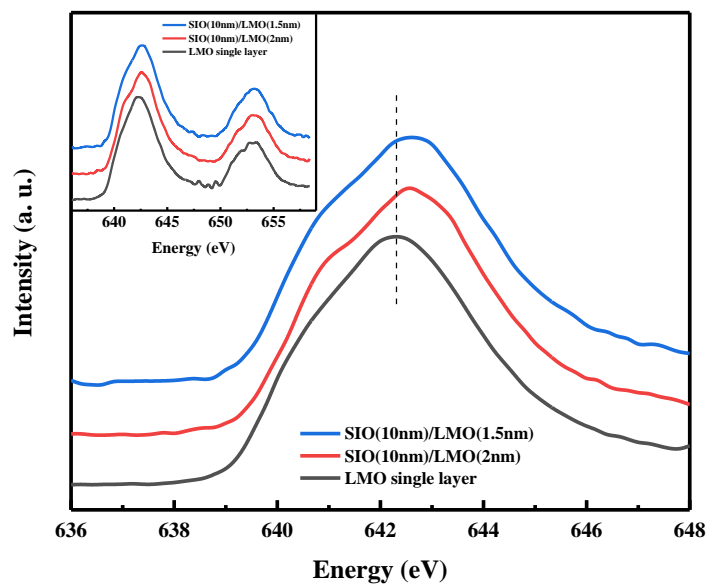

Figure 5. Mn $L_3$ normalized XAS spectra for the LMO single layer and the bilayer

with different thickness of the LMO. Inset: Mn $L_{2,3}$ normalized XAS spectra for these samples.

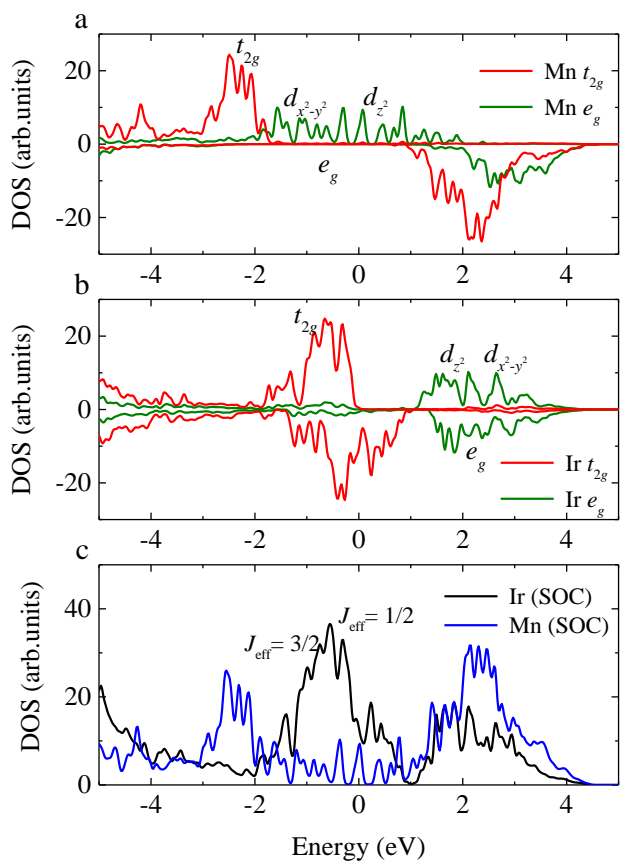

Figure 6. The spin density of states of (a) Mn and (b) Ir at the interface of $SrO/IrO_2/LaO/MnO_2$ calculated without SOC; and (c) their density of states calculated with SOC.